\documentclass[aps,pra,twocolumn,showpacs,superscriptaddress,floatfix,10pt]{revtex4-1}
\usepackage{framed}
\usepackage{graphicx}
\usepackage{amsmath}
\usepackage{epsfig}
\usepackage{helvet}
\usepackage{amssymb}
\usepackage{framed}
\usepackage{hyperref}

\begin{document}

\title{Einstein's Equations from Varying Complexity\\
}

\author{Bart{\l}omiej Czech} 
\affiliation{Institute for Advanced Study, 1 Einstein Drive, Princeton, NJ 08540}
\vskip 0.25cm
\date{\today}

\begin{abstract}
\noindent
A recent proposal equates the circuit complexity of a quantum gravity state with the gravitational action of a certain patch of spacetime. Since Einstein's equations follow from varying the action, it should be possible to derive them by varying complexity. I present such a derivation for vacuum solutions of pure Einstein gravity in three-dimensional asymptotically anti-de Sitter space. The argument relies on known facts about holography and on properties of Tensor Network Renormalization, an algorithm for coarse-graining (and optimizing) tensor networks.
\end{abstract}


\maketitle

\textit{Introduction.---} The AdS/CFT correspondence (holographic duality) \cite{holography} is the most powerful known approach to quantum gravity. It posits that every physical quantity in $d+1$-dimensional gravity with asymptotically anti-de Sitter (AdS$_{d+1}$) boundary conditions can be mapped to a corresponding quantity in a conformal field theory living on its asymptotic boundary (CFT$_d$). Although the dictionary relating AdS observables to CFT data has been studied in detail, several key aspects of AdS gravity 
have not yet been translated to the CFT language.  The present paper is concerned with a conjectured translation of one important gravitational phenomenon: that a black hole grows deeper for an exponentially long time.

Ref.~\cite{complexity} proposed that the said growth corresponds to the growing circuit complexity of the quantum state in the dual CFT. In the most recent version of the conjecture, the depth of the black hole is quantified by the gravitational action $\mathcal{A}$ inside a Wheeler-de Witt patch---the part of spacetime that is spacelike separated from a time slice of the asymptotic boundary. Living on that slice is an instantaneous quantum state of the CFT. Its complexity $\mathcal{C}$ is the minimal number of isometric gates required to assemble this state starting from some simple 
reference state. The conjecture in \cite{complexity} is that these two quantities are proportional: $\mathcal{A} \propto \mathcal{C}$.

If complexity is action, it should be possible to vary complexity and obtain Einstein's equations. The present paper reports such a derivation. A key aspect of this exercise is that I do not assume $\mathcal{A} \propto \mathcal{C}$. Instead, I work directly with the microscopic definition of circuit complexity and, to make contact with Einstein's equations, use independently known facts about AdS gravity. The present work therefore provides a novel check of the
$\mathcal{A} \propto \mathcal{C}$ conjecture. However, because circuit complexity is currently not well understood, the strategy of starting from a microscopic definition of complexity is only feasible in a restricted class of AdS geometries:

\textit{Vacuum solutions of pure Einstein gravity in 3d.---} This is a rich and varied class of spacetimes: it includes black hole solutions \cite{btz} and horizon-free geometries related to global pure AdS$_3$ by large diffeomorphisms \cite{banados}. This richness, however, is global in character; locally, all these solutions are pure AdS$_3$ because three-dimensional gravity has no propagating degrees of freedom. Thus, the content of Einstein's equations in three-dimensional pure gravity with a negative cosmological constant is to impose the locally AdS$_3$ condition. This condition can be expressed in many ways. The most convenient formulation for my purposes is in kinematic space \cite{intgeom, janmichalrob}. 

Kinematic space is the space of pairs of points in the boundary CFT. For the case at hand---an asymptotically AdS$_3$ spacetime in Lorentzian signature---the dual CFT lives on a 1+1-dimensional manifold whose lightlike coordinates will be denoted $u = x-t$ and $\bar{u} = x+t$. Kinematic space is then a four-dimensional space with coordinates $u,\bar{u},v,\bar{v}$. A function on kinematic space which characterizes the bulk geometry is the length of the bulk geodesic that connects the boundary points $(u,\bar{u})$ and $(v,\bar{v})$. 
According to the Ryu-Takayanagi proposal \cite{rt}, this length (in Planck units) equals the entanglement entropy of the CFT interval with endpoints at $(u,\bar{u})$ and $(v,\bar{v}$). 
Using the standard relation $3L/2G_N = c$ ($L$ is the AdS curvature scale and $c$ the CFT$_2$ central charge), we may write the entanglement entropy as $S_{\rm tot}(u,\bar{u},v,\bar{v}) = S(u,v) + \bar{S}(\bar{u},\bar{v})$ with:
\begin{equation}
S \! = \! \frac{c}{12}\!
\log\!\frac{\big(A(u)-A(v)\big)^2}{\delta^2 A'(u)A'(v)}
\qquad
\bar{S} \! = \! \frac{c}{12}\! 
\log\! \frac{\big(B(\bar{u})-B(\bar{v})\big)^2}{\delta^2 B'(\bar{u})B'(\bar{v})}
\label{ssbar}
\end{equation}
These objects---the left-moving and right-moving contributions to CFT entanglement entropies---obey
\begin{align}
\partial_u \partial_v \left( -\frac{6}{c} S \right) = -\frac{1}{\delta^2} \, e^{2 \left(-\frac{6}{c}S \right)} \label{ees}
\end{align}
and an identical equation for the barred quantities \cite{janmichalrob}. These two equations are the non-linear vacuum Einstein's equations in AdS$_3$, translated into the boundary language. Every real solution---parameterized by the functions $A(u)$ and $B(\bar{u})$---corresponds to a locally AdS$_3$ geometry and to a quantum state in the dual CFT \cite{banados}. My goal is to derive these equations by working with the complexity of the requisite quantum states.


\textit{States dual to locally AdS$_3$ geometries.---} 
These states comprise the CFT$_2$ ground state and its Virasoro descendants. As a first step in the argument, I shall estimate the complexity of this class of states by examining the Euclidean path integrals which prepare them. Specifically, for the ground state
\begin{equation}
\Psi(\tilde{\varphi}(x)) 
= \int e^{-S_{\rm CFT}(\varphi)}
\prod_x \prod_{\delta<z<\infty} 
D\varphi(z,x)\big|_{\varphi(\delta, x) = \tilde{\varphi}(x)}
\label{pathintegral}
\end{equation}
computes the weight of a field configuration $\tilde{\varphi}(x)$ in space while $z$ parameterizes the Euclidean time. Wave-functions of Virasoro descendants can be computed by similar formulae, but with $\tilde{\varphi}$ specified on other cutoff surfaces instead of $z = \delta$. Indeed, the Virasoro algebra (which corresponds to large diffeomorphisms of AdS$_3$) is the algebra of transformations of the cutoff surface. 

Eq.~(\ref{pathintegral}) represents one preparation of the ground state. Changing the background over which the path integral is performed away from the cutoff surface $z = \delta$ gives rise to other preparations of the same state, up to normalization. (Changing the cutoff surface would take the state around the orbit of Virasoro symmetry.) This follows from the transformation rule of the measure $D\varphi(z,x)$ under Weyl transformations \cite{gm}:
\begin{equation}
[D\varphi]_{e^{2\phi} (dx^2 + dz^2)} = 
e^{S_L[\phi] - S_L[0]} \cdot 
[D\varphi]_{(dx^2 + dz^2)}
\label{measureanomaly}
\end{equation}
Here $S_L[\phi]$ is the Liouville action
\begin{equation}
S_L = \frac{c}{24\pi} \int dx \int_\epsilon^\infty dz
\left[ (\partial_x \phi)^2 + (\partial_z \phi)^2 + 
\delta^{-2} e^{2\phi} \right]  
\label{liouvilleaction}
\end{equation}
for a field $\phi(z,x)$, which sets the Weyl frame of the Euclidean space
\begin{equation}
ds^2 = e^{2\phi} (dz^2 + dx^2) 
\equiv g_{ab}\, dx^a dx^b
\label{weylmetric}
\end{equation}
over which the path integral is carried out. The coupling constant in (\ref{liouvilleaction}) can be reabsorbed into a shift of $\phi$. I will discuss the merits of setting it to $\delta^{-2}$ below.

My strategy is to consider the Euclidean path integral performed over the background (\ref{weylmetric}) as one preparation of the quantum state. Every choice of $\phi(z,x)$ which satisfies an appropriate boundary condition gives rise to one such preparation. In the case of the ground state, the boundary condition is $\phi(\delta,x)=0$; 
for other states, we will also set the boundary condition $\phi = 0$, but on other curves through $x$-$z$ space. The objective is to characterize the complexity of the path integral $\mathcal{C}[\phi]$ as a functional of $\phi(z,x)$. Varying such a functional will then identify the minimally complex preparation of the state. 

Ref.~\cite{tadashi} proposed that the complexity of a path integral carried over background (\ref{weylmetric}) is the Liouville action shown in eq.~(\ref{liouvilleaction}). Because the justification of that claim was mostly heuristic, here I would like to offer an independent argument for why $\mathcal{C}[\phi] \propto S_L[\phi]$. To do so, I will consider the action of Weyl transformations on a discretized Euclidean path integral presented in the form of a tensor network. Recasting the problem in the language of tensor networks will have an added benefit later on.

\begin{figure}[!t]
\begin{center}
\includegraphics[width=8.5cm]{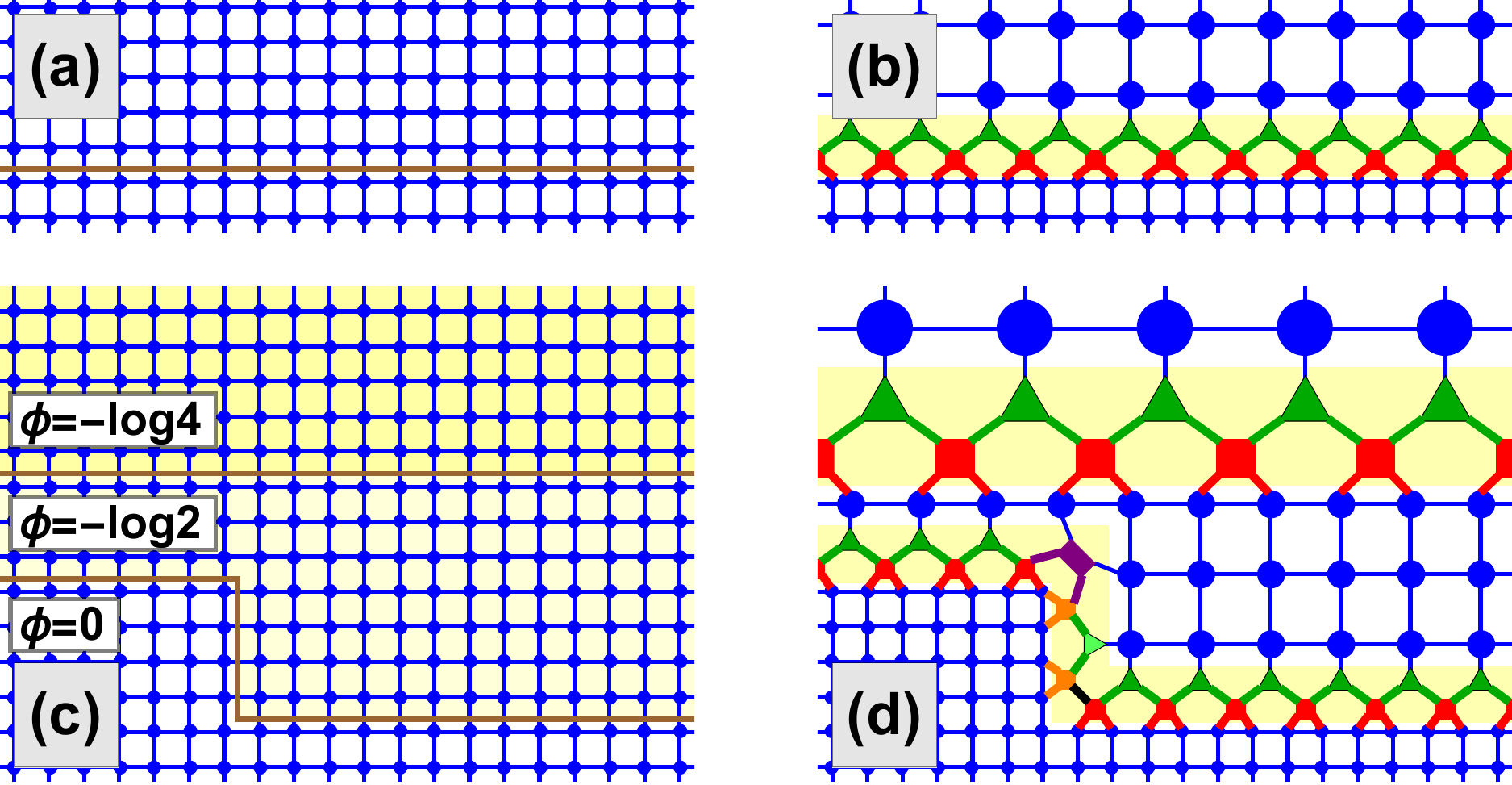}
\caption{A discretized Euclidean path integral can be represented as a regular lattice of identical tensors with nearest neighbor contractions (a). Tensor Network Renormalization approximates it with a coarser lattice (b); interfaces between the finer and coarser lattices are isometric layers (highlighted). For a piece-wise constant $\phi(z,x)$ on the lattice (c), the corresponding conformal transformation can be implemented with iterative applications of TNR (d).}
\label{fig:tnr}
\end{center}
\end{figure}

\textit{Weyl transformations of discretized path integrals.---} Before applying a Weyl transformation, the path integral (\ref{pathintegral}) over a flat half-plane is well approximated as a tensor network shown in fig.~\ref{fig:tnr}(a). We shall apply a discrete Weyl transformation to the lattice inhabited by these tensors. Following \cite{TNRlocal}, this will be done through iterative applications of the Tensor Network Renormalization (TNR) algorithm \cite{tnr}.

A single application of TNR to the discrete path integral on a regular lattice is shown in fig.~\ref{fig:tnr}(b). 
The output is a uniform network that is outwardly identical to the input network except that the density of the tensors becomes diluted by a factor of $(1/2)^2$. As a consequence of conformal symmetry, the tensors comprising the output network are the same as the tensors in the input, up to inherent ambiguities such as the choice of basis on each leg. Although the coarse-graining effected by TNR is an approximation, the difference between the states prepared by the initial and the diluted network can be made small, controlled by the bond dimension of the tensors. 
In applications to holography we imagine networks with large bond dimensions, perhaps of order $e^c$, such that the error incurred by TNR is negligible.

While TNR dilutes the interior tensors in the network, the density of external legs---the legs which span the Hilbert space where the state under construction lives---remains unaffected. Mediating between the diluted interior network and the un-diluted external legs is an extra isometric layer of tensors whose detailed form is determined by the TNR algorithm. Its purpose is to embed the coarse-grained state prepared by the diluted network in the original, fine-grained Hilbert space. An analogous isometric layer will form on every interface separating regions of the network that have been coarse-grained to varying degrees; see fig.~\ref{fig:tnr}(b).

We are now ready to apply a Weyl rescaling to a discrete path integral. If TNR is to emulate it, $\phi(z,x)$ must be a piece-wise constant function on the lattice which jumps by multiples of $\log 2$. An example discrete profile of $\phi(z,x)$ is displayed in fig.~\ref{fig:tnr}(c); recall the boundary condition $\phi = 0$. I shall apply TNR in steps. In the $k^{\rm th}$ step, I hold the legs across which $\phi(z,x)$ jumps from $-(k-1)\log2$ to $-k \log2$ fixed and coarse-grain the discrete path integral everywhere above. The region where $\phi(z,x) = -k \log2$ will eventually become diluted by a factor of $2^{-k} = e^\phi$. An example network obtained from this procedure is shown in fig.~\ref{fig:tnr}(d). For more details on using TNR to Weyl-transform path integrals, see \cite{TNRlocal}.

\textit{The complexity of the transformed path integral.---}
After the Weyl rescaling, the discrete path integral comprises two types of tensors. The first are the same tensors, which made up the initial, untransformed path integral---except that the local density of such tensors at $(z,x)$ is $e^{2\phi(z,x)}$. To account for their complexity, $\mathcal{C}[\phi]$ should include a term proportional to $\int\! dxdz\, e^{2\phi}$. 

The second component are the isometric layers. 
Because such layers follow curves where the discretized $\phi(z,x)$ jumps, accounting for their complexity will require adding to $\mathcal{C}[\phi]$ a term proportional to:
\begin{equation}
\!\!\int dx\, dz \sqrt{g}\,
g^{ab}\, \partial_a \phi\, \partial_b \phi \!=\! 
\int dx\, dz\,  
\big( (\partial_x \phi)^2 + (\partial_z \phi)^2\big)
\label{layers}
\end{equation}
This is the lowest order, rotationally symmetric expression that is even in $\vec{\nabla} \phi$. 
To see that no extra powers of $e^\phi$ are necessary, observe that a constant physical density of isometric tensors ($g^{ab} \partial_a \phi \partial_b \phi = const.$) should correspond to a density of isometries per coordinate unit length of isometric layer that goes as $e^\phi$. This is reproduced by $\sqrt{g}$ (density per unit coordinate area) times the coordinate thickness of a layer, which goes as $e^{-\phi}$. 

Assuming the average complexity of isometric layers (per unit area) is greater than the complexity of the tensors of the first type, accounting for the isometries demands including in $\mathcal{C}[\phi]$ a positive multiple of (\ref{layers}). (If the isometries were less complex than the initial tensors, they would require supplementing $\int \! dx dz\, e^{2\phi}$ with a negative multiple of (\ref{layers}).) Combining both terms, we get $\mathcal{C}[\phi] \propto S_L[\phi]$. For now, their relative coefficient can be readjusted by a shift of $\phi$, but the choice in (\ref{liouvilleaction}) will prove meaningful and convenient below. 

\textit{The optimal path integral: state complexity.---} 
To find the complexity of the state, we seek the least complex circuit that prepares the state. Setting the variation of expression (\ref{liouvilleaction}) to zero, we obtain the equation of motion
\begin{equation}
4 \partial_w \partial_{\bar{w}} \phi = \delta^{-2} e^{2\phi},
\label{liouvilleeom}
\end{equation}
where $w = x + iz$ and $\bar{w} = x - iz$. Evaluating `action' (\ref{liouvilleaction}) on a solution of (\ref{liouvilleeom}) yields the complexity of a quantum state, which is specified by the boundary condition of $\phi$. For the simplest boundary condition $\phi(\delta,x)=0$ that selects the CFT ground state, we get $\phi = -\log (z/\delta)$. 

\begin{figure}[!t]
\begin{center}
\includegraphics[width=8.5cm]{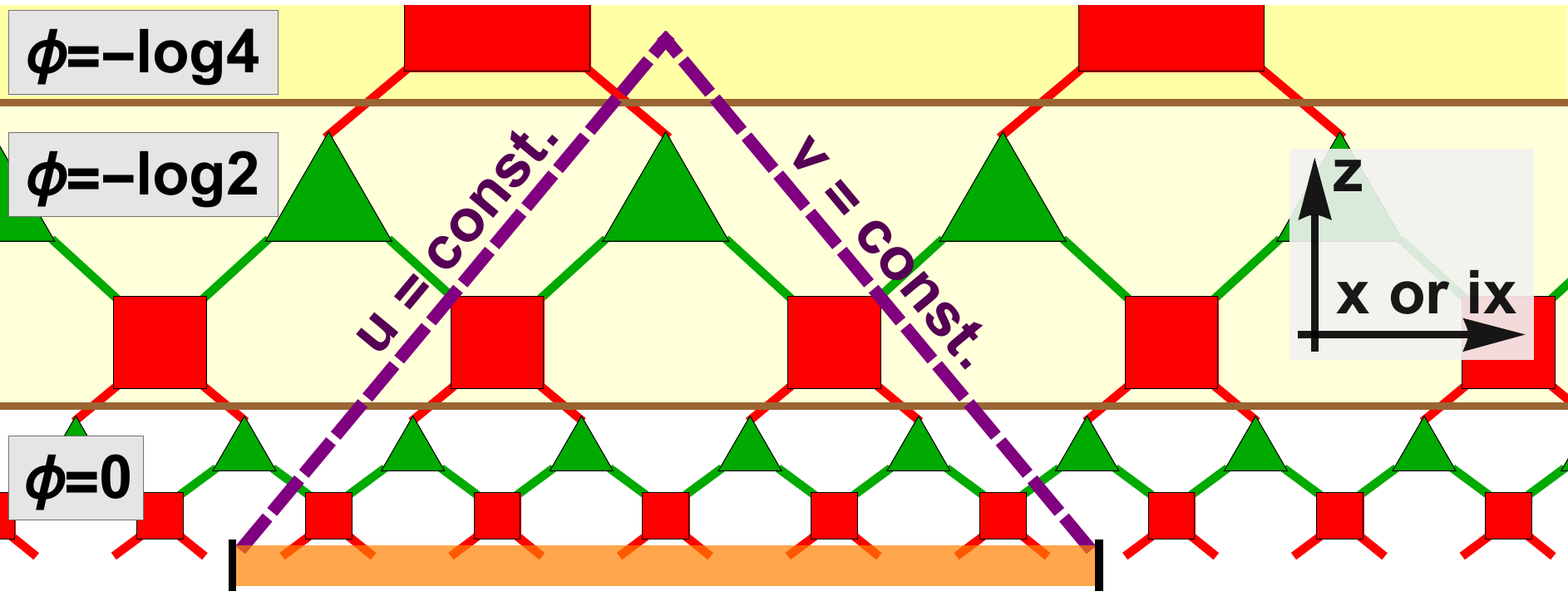}
\caption{The optimal network (MERA) consists entirely of isometric layers. The entanglement entropy of interval $(u,v)$ can be approximated by counting the isometric layers through which the `exclusive causal cone' of the interval passes. The layers are indexed by $\phi/(-\log 2)$.
}
\label{fig:mera}
\end{center}
\end{figure}

The remaining task is to connect equation (\ref{liouvilleeom}) to (\ref{ees}). Before doing so, let us pause for a useful observation:

\textit{The optimal network is MERA.---} 
In discrete settings, eq.~(\ref{liouvilleeom}) demands a `greedy' coarse-graining: a sequence of consecutive isometric maps, without retaining any diluted tensors from the initial path integral.
Heuristically, keeping the non-isometric tensors has no advantage: they cost complexity but do not expedite preparing the state because with or without them the same isometries must still be applied. 
The optimal network, shown in fig.~\ref{fig:mera}, 
is known as MERA \cite{mera}; the fact that `greedy' iterations of TNR produce MERA was first observed in \cite{tnr2mera}. The present argument suggests that MERA or a close analogue is the most efficient circuit for preparing the CFT$_2$ ground state and Virasoro descendants. In the continuum, a likely candidate for optimality is cMERA \cite{cmera}, its continuous version. Identifying MERA as the most efficient circuit reproduces the intuitions of \cite{complexity, tadashi, tadashi0}. 

One bonus of the preceding discussion is that MERA is made up of unitary and isometric tensors. A possible objection to quantifying state complexity using Euclidean path integrals is that the latter may not be prepared by a sequence of isometric gates. In contrast, most discussions of circuit complexity assume the elementary gates comprise only unitaries and isometries, which seems to disallow the tensor networks shown in fig.~\ref{fig:tnr}. Since MERA involves only unitaries and isometries, this objection does not apply. If MERA is optimal 
in a larger class of circuits that includes Euclidean path integrals then it is also optimal in the narrower class of isometric circuits.

A more serious danger is that TNR is not versatile enough to study state complexity. It is possible the optimal circuit cannot be reached by applying TNR to the path integral. The only response to this valid objection is that proving optimality in quantum field theory is nearly impossible and will remain so in the foreseeable future, so a pedantic insistence on verifying optimality would postpone any practical inspection of circuit complexity {\it ad kalendas Graecas}. Having duly noted the logical possibility that MERA is not optimal, 
I proceed to discuss:

\textit{The Lorentzian geometry of the ground state MERA.---} 
The conclusion thus far is that the optimal circuit (MERA) lives on a Euclidean geometry with a metric:
\begin{equation}
ds^2 = -\frac{4\delta^2 A'(w) B'(\bar{w})}{(A(w)-B(\bar{w}))^2}
\,dw \,d\bar{w}\,.
\label{euclideanmetric}
\end{equation}
The factor in front of $dw d\bar{w}$ is a general solution of eq.~(\ref{liouvilleeom}) for $e^{2\phi}$. However, the MERA network is also known to have a causal structure (see fig.~\ref{fig:mera}), which is a consequence of the isometric and unitary character of its tensors \cite{mera, beny, TNfromKS}. Can we use eq.~(\ref{euclideanmetric}) to understand the Lorentzian manifold that captures MERA's causal structure?

It is most instructive to do so first for the ground state, 
for which $A(w)=w$ and $B(\bar{w})=\bar{w}$. Eq.~(\ref{euclideanmetric}) now reads
\begin{equation}
ds^2 = 
-\frac{4\delta^2\, dw\, d\bar{w}}
{(w-\bar{w})^2} = 
\frac{dx^2+dz^2}{(z/\delta)^2}
\label{canonh2}
\end{equation}
and $\phi(z,x) = -\log(z/\delta)$. Recalling the way TNR produced MERA, we recognize that every successive jump of $\phi$ by $\log2$ marks one additional isometric layer. The causal structure of MERA, which tracks which tensors impact the state on which external legs, indicates that a tensor at $x_0$ and $z = 2^k\delta$, i.e. on the $k^{\rm th}$ layer of MERA, impacts the state on legs $x_0 - 2^k\delta$ through $x_0 + 2^k\delta$. 
Thus, the lightlike directions on the Lorentzian manifold underlying MERA are simply $x \pm z$. Meanwhile, its volume form (in the discrete language, the number of tensors) is the same as in eq.~(\ref{canonh2}). Altogether, we conclude that the Lorentzian geometry of MERA is encapsulated by:
\begin{equation}
ds^2 = (-dx^2 + dz^2)\big/(z/\delta)^2\,.
\label{lorentzianmetric}
\end{equation}
This reasoning was spelled out before e.g. in \cite{beny, TNfromKS}; see also \cite{tadashi0} for a comparison of how the Lorentzian and Euclidean geometries of MERA can be embedded in AdS$_3$. Note that we could have obtained metric~(\ref{lorentzianmetric}) directly from (\ref{euclideanmetric}) by continuing $x \to ix$ and
\begin{equation*}
w\! =\! x + iz \!\to\! i(x+z)\! \equiv\! iv
\qquad\,\, 
\bar{w}\! =\! x - iz \!\to\! i(x-z) \!\equiv\! iu 
\end{equation*}
so that eq.~(\ref{canonh2}) becomes
$- 4\delta^2\, dv\, du / 
(v-u)^2$.

\textit{Ground state $\phi$ is minus entanglement entropy.---} 
In holographic interpretations of MERA such as \cite{swingle, TNfromKS}, one assumes that the entanglement entropy of an interval can be estimated by counting the legs which cross its `exclusive causal cone' (see fig.~\ref{fig:mera}). For an interval $(u=x_0 - 2^k\delta, v=x_0 + 2^k\delta)$, this number is $2k$---two legs (on left and right) for each MERA layer from the UV up to the top of the causal cone. In terms of the lightcone coordinates $u$ and $v$, the entanglement entropy of interval $(u,v)$ therefore equals:
\begin{equation}
\frac{S_{\rm tot}(u,u,v,v)}{\# \cdot c} 
= -\frac{2\phi(w=iv,\bar{w}=iu)}{\log 2}.
\label{identif}
\end{equation}
Note that the entanglement entropies of $(2\delta)$-sized intervals correctly vanish; if we had reabsorbed the coupling constant in (\ref{liouvilleaction}) into a shift of $\phi$, we would have had to undo the shift in (\ref{identif}) with the same physical outcome. 
Quantity $\# \cdot c$ is the additive contribution to the entanglement entropy from every line crossed by a minimal cut; the notation emphasizes that it is an $\mathcal{O}(1)$ multiple of the central charge. To reconcile eq.~(\ref{ssbar}) with $\phi(z,x) = -\log(z/\delta)$ for the vacuum, the constant $\#$ must be $(\log2)/6$, which yields for $S(u,v)$ and $\bar{S}(\bar{u}, \bar{v})$:
\begin{align}
-\frac{6}{c}S(u,v) & = \phi(w = iv, \bar{w}=iu)
\label{finalident} \\
-\frac{6}{c}\bar{S}(-\bar{v}, -\bar{u}) & = \phi(w = -i\bar{u}, \bar{w}=-i\bar{v}) \nonumber
\end{align}
The step from (\ref{identif}) to (\ref{finalident}) is motivated by considering both analytic continuations $x \to \pm ix$ of $\phi(w,\bar{w})$ and noting that they transform into one another under spatial reversal, as do $S(u,v) \leftrightarrow \bar{S}(-\bar{v}, -\bar{u})$.  Written in terms of $u$ and $v$, eq.~(\ref{liouvilleeom}) is synonymous with (\ref{ees}) and its barred counterpart.

\textit{Time reversal-invariant Virasoro descendants.---} Consider general functions 
present in (\ref{euclideanmetric}) as coordinate changes ($W = A(w)$ and ($\overline{W} = B(\bar{w})$) and repeat the argument. Metric~(\ref{euclideanmetric}) reduces to the canonical form (\ref{canonh2}) with $w,\bar{w} \to W, \overline{W}$ and $\phi=-\log({\rm Im}(W)/\delta)$ again indexes consecutive layers of the optimal (MERA) network. Its causal structure is captured by continuing $W \to iV \equiv A(iv)$, which defines the Lorentzian metric
\begin{equation}
\!\!\!ds^2 
\!= \!
\frac{4\delta^2 A'(iv) B'(iu)\, du\, dv}{(A(iv)-B(iu))^2} 
\!= \!
-e^{2\phi(w=iv, \bar{w}=iu)} du\, dv.
\end{equation}
If identification (\ref{finalident}) still holds, eq.~(\ref{liouvilleeom}) that $\phi(w,\bar{w})$ obeys will again reduce to (\ref{ees}).

There are several ways to confirm identification~(\ref{finalident}). Combining it with the equation of motion (\ref{liouvilleeom}), we see that the Lorentzian geometry of MERA is
\begin{equation}
-\!e^{2\phi(w=iv, \bar{w}=iu)} du\, dv 
= (-24\delta^2/c)\, \partial_{u} \partial_{v}
S(u,v)\, du\, dv,
\label{ksmetric}
\end{equation}
i.e. kinematic space \cite{intgeom, janmichalrob}. The same conclusion was reached in \cite{TNfromKS} for independent reasons, which involved a tensor network interpretation of the differential entropy formula \cite{diffent}. Yet another argument observes that both sides of (\ref{ksmetric}) are reparameterization-invariant, so if (\ref{finalident}) holds in the $W, \overline{W}$ coordinates, it must hold in $w,\bar{w}$ too.

\textit{General Virasoro descendants.---} 
This reasoning applies to time reversal-invariant states, which can be prepared by real Euclidean path integrals. To extend it to arbitrary Virasoro descendants, one might complexify metric (\ref{weylmetric}) by declaring $w$ and $\bar{w}$ to be independent: 
\begin{equation}
w = x_R - it_R \qquad {\rm and} \qquad \bar{w} = x_L - it_L.
\end{equation}
If $\phi(w,\bar{w})$ still satisfies eq.~(\ref{liouvilleeom}), identification~(\ref{finalident}) would again imply eq.~(\ref{ees}). But at present TNR cannot be used to justify eq.~(\ref{liouvilleeom}) because its application to complexified path integrals remains unexplored.

\textit{Future directions.---} The next goal should be to introduce bulk matter. Other than a point particle (a conical defect), which can be done along the lines of \cite{tadashi}, this will likely require qualitatively new ingredients. Another important goal is higher dimensions. If we consider Weyl transformations of the path integral, an identical argument suggests a `complexity-action'
\begin{equation}
\mathcal{C}[\phi] \propto \int d^dx
\left[ e^{(d-2)\phi} (\partial \phi)^2 + 
\delta^{-2} e^{d\phi} \right],
\end{equation}
an expression previously conjectured in \cite{tadashi}. 
We recognize this complexity functional as pitting 
curvature against volume, with the curvature term quantifying the complexity cost of coarse-graining tensors. The optimal geometry would again be a manifold of constant curvature whose magnitude is set by the cutoff. 

\textit{Acknowledgements.---}
I thank Vijay Balasubramanian, Nele Callebaut, Xi Dong, Lampros Lamprou, Juan Maldacena, Samuel McCandlish, Rob Myers, Onkar Parrikar, Charles Rabideau, Douglas Stanford, James Sully, Tadashi Takayanagi, Eric Verlinde, Herman Verlinde, Guifr{\'e} Vidal, Manus Visser, Aron Wall and Edward Witten for useful discussions. I thank the organizers of ``Tensor Networks for Quantum Field Theories II'' held at Perimeter Institute for Theoretical Physics (supported by the Government of Canada through Industry Canada and by the Province of Ontario through the Ministry of Economic Development \& Innovation) where much of this work was discussed and completed. My research is supported by the Peter Svennilson Membership in the Institute for Advanced Study.

\end{document}